\documentclass[twocolumn,aps,superscriptaddress,nofootinbib]{revtex4-1}
\usepackage{graphicx}
\usepackage{braket} 
\usepackage{float}%
\usepackage{amsmath,amssymb}
\usepackage{bbold} 
\usepackage[normalem]{ulem} 
\usepackage[utf8]{inputenc}

\usepackage[usenames, dvipsnames]{color}
 \usepackage{xcolor}

\usepackage{bbm}
\usepackage{bbold}

\usepackage{graphicx}
\usepackage[hidelinks=true]{hyperref}
\hypersetup{
  colorlinks   = true, 
  urlcolor     = blue, 
  linkcolor    = blue, 
  citecolor   = red 
}

\def\slashchar#1{\setbox0=\hbox{$#1$}
   \dimen0=\wd0 \setbox1=\hbox{/} \dimen1=\wd1
   \ifdim\dimen0>\dimen1 \rlap{\hbox to \dimen0{\hfil/\hfil}} #1
   \else  \rlap{\hbox to \dimen1{\hfil$#1$\hfil}} / \fi}

\def\p1{\slashchar{p^\prime}}

\newcommand{\reales}{\mbox{R}\hspace{-1.0ex}\rule{0.1mm}{1.5ex}\hspace{1ex}}

\graphicspath{{./plots/}}

\begin{document}

\title{Weak Kaon Production off the nucleon and Watson's theorem}

\author{E. \surname{Sa\'ul-Sala}}
\affiliation{Departamento de F\'\i sica Te\'orica and IFIC, Centro Mixto
Universidad de Valencia-CSIC, Institutos de Investigaci\'on de
Paterna, E-46071 Valencia, Spain} 

\author{J. E. \surname{Sobczyk}}
\affiliation{Institut f\"ur Kernphysik and PRISMA$^+$ Cluster of Excellence, \\
Johannes Gutenberg-Universit\"at, 55128 Mainz, Germany} 

\author{M. Rafi \surname{Alam}}
\affiliation{Department of Physics, Aligarh Muslim University, Aligarh-202 002, India}

\author{L. \surname{Alvarez-Ruso}}
\affiliation{Departamento de F\'\i sica Te\'orica and IFIC, Centro Mixto
Universidad de Valencia-CSIC, Institutos de Investigaci\'on de
Paterna, E-46071 Valencia, Spain} 

\author{J. \surname{Nieves}}
\affiliation{Departamento de F\'\i sica Te\'orica and IFIC, Centro Mixto
Universidad de Valencia-CSIC, Institutos de Investigaci\'on de
Paterna, E-46071 Valencia, Spain} 

\begin{abstract} 
We have improved the tree-level model of Ref~\cite{RafiAlam:2010kf}  for weak production of kaons off nucleons by partially restoring unitarity. This is achieved by imposing Watson's theorem to the dominant vector and axial-vector contributions in appropriate angular momentum and isospin quantum number sectors. The observable consequences of this procedure are investigated.
\end{abstract}
\maketitle 

\section{Introduction}
A good understanding and realistic modeling of neutrino cross sections is important to reduce systematic uncertainties in oscillation experiments~\cite{Mahn:2018mai,Alvarez-Ruso:2017oui,Katori:2016yel,Alvarez-Ruso:2014bla,Formaggio:2013kya}. Much attention has been paid to quasi-elastic scattering and weak pion production,  which give a large contribution in the few-GeV neutrino energy region probed in most accelerator experiments. On the other hand, with better statistics and higher precision goals, other, largely unexplored, processes with smaller cross sections may play a significant role. Kaon, and strangeness production in general, belongs to this category.    

 The charged-kaon production ($\nu_\mu \mathrm{CH} \rightarrow \mu^- K^+ X$) measurement at MINERvA~\cite{Marshall:2016rrn} experiment opens a new window to study the weak strangeness production mechanisms in detail.  The weak processes that could lead to kaons in the final state are either initiated by strangeness conserving ($\Delta S=0$) or strangeness changing ($\Delta S=1$) mechanisms.  Although the  $\Delta S=1$ reactions ($1K$) are Cabibbo suppressed  compared to $\Delta S=0$ ones ($YK$), the latter involve the production of massive strange hyperons ($Y$), which pushes the reaction thresholds  higher in neutrino energies. Therefore, below 2 GeV of incoming neutrino energies, the 1$K$ reaction is favoured~\cite{Marshall:2016rrn,RafiAlam:2010kf}. In nuclei, final state interactions of the produced kaon are not very strong because of the absence of baryon resonances. However, kaons can also be produced in secondary collisions, rendering the extraction of information about the elementary 1$K$-production amplitudes in experiments with nuclear targets rather difficult~\cite{Lalakulich:2012gm}. As for several other processes, progress in our understanding of weak kaon production would greatly benefit from modern cross section measurements on hydrogen and/or deuterium \cite{Alvarez-Ruso:2017oui}.
 
 Theoretical work on weak production of meson-baryon pairs with open and hidden strangeness was performed in the early days of neutrino physics~\cite{Shrock:1975an,Mecklenburg:1976pk,Amer:1977fy,Dewan:1981ab} and resumed only recently with studies in the $\Delta S = 0$~\cite{Adera:2010zz,Nakamura:2015rta}, $\Delta S = -1$~\cite{Alam:2011xq,Ren:2015bsa} and $\Delta S = 1$~\cite{RafiAlam:2010kf} sectors. The first calculation of the $\nu_l N \rightarrow l^- K N'$ amplitudes using leading-order SU(3) chiral perturbation theory was performed by Alam et. al.~\cite{RafiAlam:2010kf}. The threshold cross section was predicted in a model independent way in terms of only three precisely-known quantities $f_\pi$, $D$ and $F$, where $F$ and $D$ are the couplings that appear from the SU(3) Wigner–Eckart theorem of the axial-vector current. To extend the validity of the study to higher energies, the hadronic currents were multiplied by a phenomenological global dipole form factor. However, as it is based on tree-level diagrams, this model neither respects the unitarity of the $S$ matrix, nor it satisfies the related Watson's theorem~\cite{Watson:1952ji}~\footnote{A consequence of unitarity of $S-$matrix and time reversal symmetry.}, according to which, the phase of the amplitude is determined by the strong meson-baryon  interaction ($KN$ in this case).       
 
 In the present work, we address this issue and partially restore unitarity  by imposing Watson's theorem. This is achieved by introducing relative phases in the amplitudes derived in Ref.~\cite{RafiAlam:2010kf}, as suggested by Olsson in \cite{Olsson:1974sw} for pion photoproduction. In Refs.~\cite{Alvarez-Ruso:2015eva,Hernandez:2016yfb}, the same strategy has been successfully applied to the weak pion production model of Ref.~\cite{Hernandez:2007qq}. In the following we briefly present the model for $\Delta S =1$  $K$-production and the Watson's prescription to approximately restore unitarity, followed by a discussion on the impact of this improvement on observable quantities.

\section{Formalism}\label{sec:formalism}
The allowed neutrino-induced $\Delta S=1$ single-kaon production reaction channels on nucleons are
\begin{align}\label{eq:process}
    \nu_l + p &\rightarrow  l^- + p + K^+ \nonumber \\
    \nu_l + n &\rightarrow  l^- + p + K^0 \\
    \nu_l + n &\rightarrow  l^- + n + K^+ . \nonumber
\end{align}
The differential cross section for the processes of Eq.~\eqref{eq:process} is given by
\begin{equation}\label{eq:diff_phsp}
    \frac{d^4\sigma}{ dW\, dQ^2 d\Omega_K^*} = \frac{ |\vec{p}_K^{\,*}|}
    {64 (2\pi)^4 E_\nu^2 M_N^2 } |\overline{\mathcal{M}}|^2 
\end{equation}
with
\begin{equation}
\label{eq:Msqr}
|\overline{\mathcal{M}}|^2  = \frac{1}{4} G_F^2 |V_{us}|^2 L^{\mu \nu} J_{\mu \nu} \,, 
\end{equation}
where $L^{\mu \nu}$ ($J_{\mu \nu}$) is the leptonic (hadronic) tensor; $W$ is the invariant mass of the outgoing kaon-nucleon pair while $Q^2=-q^2$ stands for minus the square of the four momentum transfer $q= k - k^\prime$, with $k$ and $k^\prime$ the four momenta of the incoming neutrino and outgoing lepton respectively.
We fix the lepton kinematics and target nucleon in the Laboratory frame, in which $E_\nu$ denotes the incoming neutrino energy $(=k^0)$. The outgoing $KN$ system is treated in the rest frame of the pair, referred to as the hadronic center-of-mass (HCM) frame. We represent HCM quantities with a  `$\ast$' superscript.  In Eq.~\eqref{eq:diff_phsp}, the kaon momentum $(\vec{p}_K^{\, *})$ and solid-angle ($\Omega_K^\ast$) are indeed in the HCM frame. 
The Fermi coupling constant ($G_F$) and the Cabibbo-Kobayashi-Maskawa (CKM) matrix element, $|V_{us}|$,  have numerical values of $1.166\times 10^{-5}$ GeV$^{-2}$ and $0.2243$ respectively~\cite{Tanabashi:2018oca}. 

\begin{figure}[h!]
\begin{center}
\includegraphics[width=0.48\textwidth]{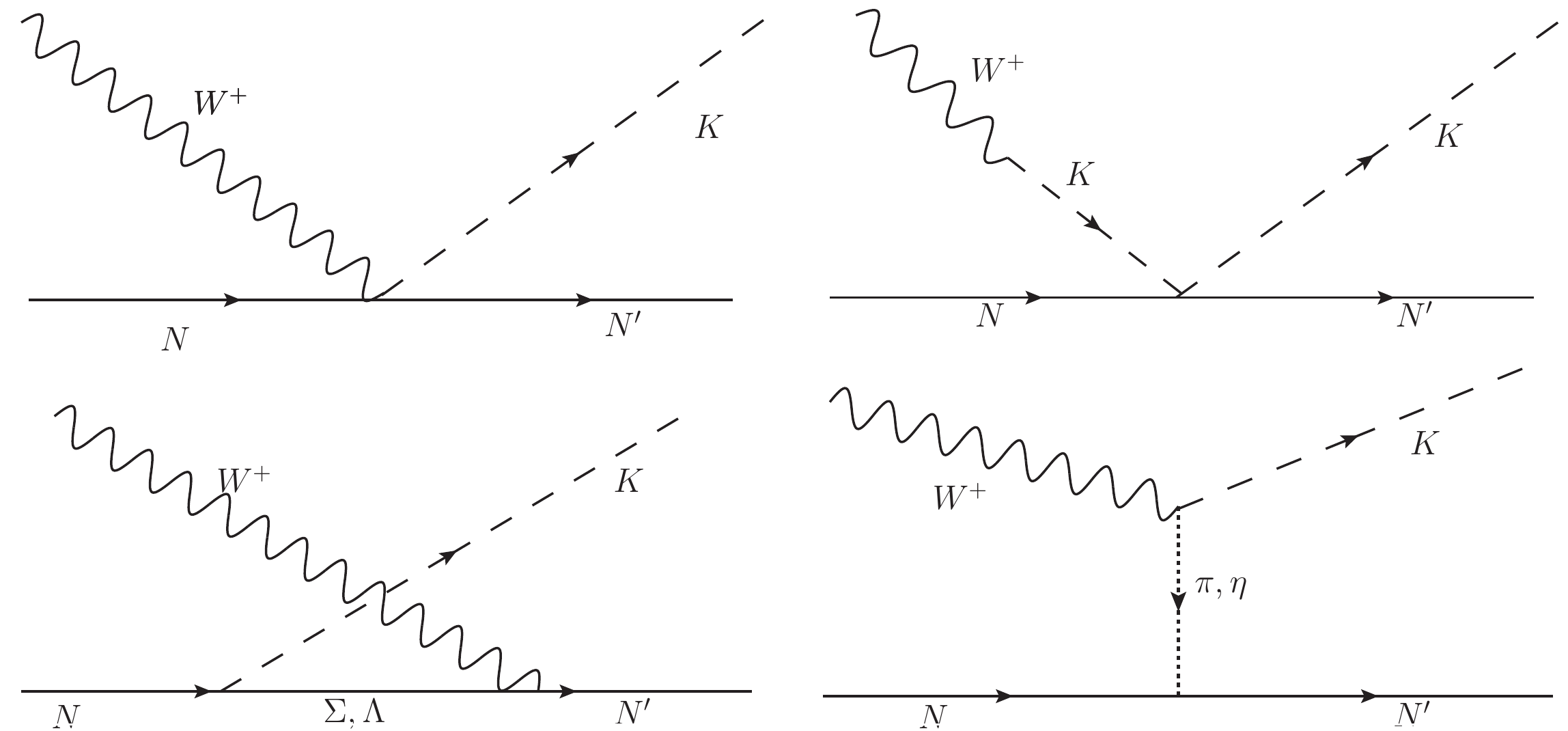}
\caption{\label{fig:diags} Feynman diagrams for the hadronic current $W^+ N\to  K N'$. From the upper left corner in clockwise order: contact (CT), kaon pole  (KP),
  $\pi$ and $\eta$ in flight ($\pi$P, $\eta$P) and $u-$channel hyperon exchange (Cr$\Sigma$, Cr$\Lambda$) terms.}
\end{center}
\end{figure}

The leptonic tensor may be written as, 
\begin{align}
    L_{\mu\nu} &= 8\left[k'_\mu \, k_\nu + k'_\nu \, k_\mu - g_{\mu\nu} (k'\cdot k) + i \epsilon^{\mu\nu\sigma\rho} k'_\sigma k_\rho \right] \,,
\end{align}
where we follow the convention $\epsilon^{0123} = +1$ for the 4-dimensional Levi-Civita tensor.  
Finally, the tensor $J^{\mu \nu}$ can be expressed in terms of the $W^+ N \rightarrow K N'$ hadronic current $j^\mu$ as
\begin{equation}
J^{\mu \nu} = \sum_\mathrm{spins} j^\mu \left(j^\nu\right)^\dagger \,,   
\end{equation}
where the sum is performed over the spin projections of the incoming and outgoing nucleons; $W^+$ denotes the virtual weak gauge boson. This hadronic current, obtained from the expansion of the SU(3) chiral Lagrangian at its lowest order, plus next-to-leading contributions to weak magnetism, was derived in Ref.~\cite{RafiAlam:2010kf}.  The complete set of diagrams that contribute to Eq.~\eqref{eq:process} are shown in Fig.~\ref{fig:diags}.
The corresponding expressions that add to $j^\mu$ are given in Eq.~(15) of Ref.~\cite{RafiAlam:2010kf}. The parameters that enter the current are well known: the pion decay constant($f_\pi$), couplings $D$ and $F$, fixed from nucleon and hyperon semileptonic decays, and measured values of nucleon magnetic moments. We refer the reader to  Ref.~\cite{RafiAlam:2010kf} for details. Finally, to extend the kinematic range of the calculation, a global dipole form factor has been introduced, with a dipole mass of $1 \pm 0.1$~GeV,  accounting for higher-order hadronic structure and its uncertainty.

\subsection{Watson's theorem for weak $K$-production}\label{sec:watson_kaon}

Let us consider matrix elements of the transition ($T$) scattering operator between two-body states with well defined total angular momentum $J$ and particle helicities ($\lambda$) in the HCM frame.\footnote{We warn the reader that, although the HCM frame is used throughout \ref{sec:watson_kaon}, we have dropped the '*' superscript to maintain the readability of equations.} Following the derivation of Sec. II.A of  Ref.~\cite{Alvarez-Ruso:2015eva} for weak pion production, the $S-$matrix unitarity and time reversal symmetry imply that   
\begin{align}
& \sum_{\lambda_{K''}\lambda_{N''}} \langle J, M; \lambda_{K''},\lambda_{N''}\,|T(s)|J, M; \lambda_K,\lambda_{N'}\, \rangle^*    \nonumber \\
& \times \langle J, M;\lambda_{K''},\lambda_{N''}\,|T(s)|J, M; \lambda_W,\lambda_{N}\,\rangle
\in \reales \,, \label{eq:watJM}
\end{align}
for the $W^+ N \rightarrow K N'$ transition. In the present study, the center-of-mass energy of the kaon-nucleon system, $\sqrt{s}=W$, is limited to the range in which the only relevant intermediate states in Eq.~\eqref{eq:watJM} are $K''N''$ pairs.  Therefore, this equation, Watson's theorem, relates the phases of the strong $K'' N'' \rightarrow K N'$ amplitudes with the electroweak $W N \rightarrow K'' N''$ ones. The later, up to a real normalization constant 
\begin{equation}
\langle K'' N'' | T | W N  \rangle \propto  -i j_\mu \epsilon^\mu \,,   
\end{equation}
in terms of the hadronic current $j^\mu$ introduced above and the polarization vector of the $W$ boson.\footnote{Notice that the gauge coupling has been factored out and absorbed in the Fermi constant of Eq.~(\ref{eq:Msqr}).} The $W$-boson offshellness does not affect the present argument~\cite{Alvarez-Ruso:2015eva}. As stated above, we consider only $KN$ intermediate states in Eq.~\eqref{eq:watJM}, restricting the validity of the approach to invariant masses of the $KN$ pair below the $K K Y$ threshold. We further neglect the influence of $K\pi N$ intermediate states. This assumption relies on the observation that in the $KN$ partial waves under consideration (details are given below), inelasticities are either sharply or very close to one for invariant masses below 2.1~GeV~\cite{SAID}. 

To be more specific, in Eq.~\eqref{eq:watJM} after setting  the kaon helicities to zero, we denote as $r$ the  helicity of the $W$ gauge boson,  and as $\lambda, \lambda', \rho$ the corresponding ones of the initial, final and intermediate nucleons. Furthermore, assigning the $z$ direction ($\theta=\varphi=0$) to the $WN$ incoming pair, one can write 
\begin{equation}
\ket{\theta=0,\varphi=0; r,\lambda} 
= \sum_J \sqrt{\frac{2J+1}{4\pi}} \ket{J,M=r - \lambda; r\, \lambda}
\label{eq::states}
\end{equation}
which follows from Eq.~\eqref{eq::relation_between_states_noCJ} of Appendix~\ref{app:states}. By taking into account that $T$ is a scalar and therefore diagonal in $J$, Eq.~\eqref{eq:watJM} can be cast as
\begin{align}
\sum_{\rho}
& \langle J,M;\underbrace{0,\rho}_{K N}|T(s)|J,M;\underbrace{0,\lambda'}_{K N}
\rangle^*   \nonumber \\
& \times \langle J,M;\underbrace{0,\rho}_{K N}|T(s)|\theta,\varphi=0;\underbrace{r,\lambda}_{W
N}\rangle \in\reales\,, \label{eq:watJM1}
\end{align}
 with $M= r -\lambda$. Introducing states with well-defined orbital angular momentum $L$ and spin $S$, and using their transformation properties given in Appendix~\ref{app:states}, one finds 
 \begin{align}
& \sum_{L}\sum_{\rho}\frac{2L+1}{2J+1} (L,1/2,J|0,-\lambda',-\lambda')
(L,1/2,J|0,-\rho,-\rho) 
\nonumber\\
& \times\underbrace{\langle J,M;{L,1/2}|T(s)|J,M;{L,1/2}
\rangle^*}_{K N\to K N}
\nonumber\\
& \times \underbrace{\langle
  J,M;{0,\rho}|T(s)|\theta,\varphi=0;{r,\lambda}\rangle}_{WN\to K N}
\in\reales \,, \label{eq:watson-gral}
\end{align}
given that parity is conserved by the strong $K N \rightarrow K N$ amplitudes. Here $(L,S,J|M_L,M_S,M_J)$ are Clebsch-Gordan coefficients. 

Based on the behavior of weak kaon production amplitudes close to threshold, 
it is reasonable to assume that the process under study is dominated by the $s-$partial wave ($L=0$). This implies that $S=J=1/2$, the nucleon spin. Equation~\eqref{eq:watson-gral} takes then the form
\begin{equation}
\chi_{r,\lambda}(s) \langle 1/2,r-\lambda;{0,1/2}|T(s)|1/2,r-\lambda;{0,1/2} \rangle^*   \in\reales \end{equation}
where the shorthand notation
\begin{align}\label{eq:chi1}
\chi_{r,\lambda}(s)= 
\sum_{\rho} \,\langle
1/2,r-\lambda; {0,\rho}|T(s)|\theta,\varphi=0;{r,\lambda}\rangle
\end{align}
has been introduced. Up to an irrelevant constant, these functions can be written as
\begin{widetext}
\begin{equation}\label{eq:chi_integrated}
 \chi_{r,\lambda}(s)
     =  \sum_{\rho} \int d\Omega\ {\cal D}^{(1/2)}_{M\ -\rho}(\varphi,\theta,-\varphi)   \braket{\theta, \varphi; 0, \rho | T(s) | \theta,\varphi=0; r, \lambda}    
\end{equation}
where ${\cal D}^{(1/2)}_{M\ -\rho}$ are Wigner D-matrices [see Eq.~\eqref{eq::relation_between_states_noCJ} in Appendix \ref{app:states}]. The integral is performed over the solid angle of the outgoing kaon in the HCM frame.
\end{widetext}

Owing to the $V - A$ nature of the weak interaction, $T$ in Eq.~\eqref{eq:chi1} can be expressed as $T_V - T_A$, $T_{V(A)}$ being even (odd) under parity inversion. Therefore, it is convenient to write $\chi_{r,\lambda} = \chi_{r,\lambda}^V - \chi_{r,\lambda}^A$.
We then explore the transformation properties of $\chi_{r,\lambda}(s)$ under parity from which the following relations are deduced (see  Appendix \ref{app::parity}): 
\begin{equation}\label{eq::chi_V_A}
\begin{aligned}
    \chi_{r,\lambda}^V &= \frac{1}{2} \left( \chi_{r,\lambda} - \chi_{-r,-\lambda} \right)\,, \\ 
    \chi_{r,\lambda}^A &= - \frac{1}{2} \left( \chi_{r,\lambda} + \chi_{-r,-\lambda} \right) \,.
\end{aligned}
\end{equation}
They allow to reduce the number of independent functions from four vector (axial) ones to two \cite{Alvarez-Ruso:2015eva} for each of the reaction channels listed in Eq.~\eqref{eq:process}.\footnote{Combinations with $|r-\lambda|=3/2$ are excluded because $J=1/2$.} 

Finally, we project onto states with well defined isospin ($I$), introducing isospin amplitudes, and the corresponding $\chi^{(I=0,1)}$ functions 
\begin{equation}\label{eq::chi_Iso}
    \begin{aligned}
    \chi^{(1)} &= \chi (W^+ \, p \rightarrow K^+ \, p) \,, \\
    \chi^{(0)} &= \chi (W^+ \, n \rightarrow K^+ \, n) - \chi (W^+ \, n \rightarrow K^0 \, p) \,.
    \end{aligned}
\end{equation}
Other indices have been dropped for simplicity. These identities allow us to write the $\chi$ functions  for all three processes in terms of only two with $I=0,1$.

\begin{figure*} 
\includegraphics[width=0.45\textwidth]{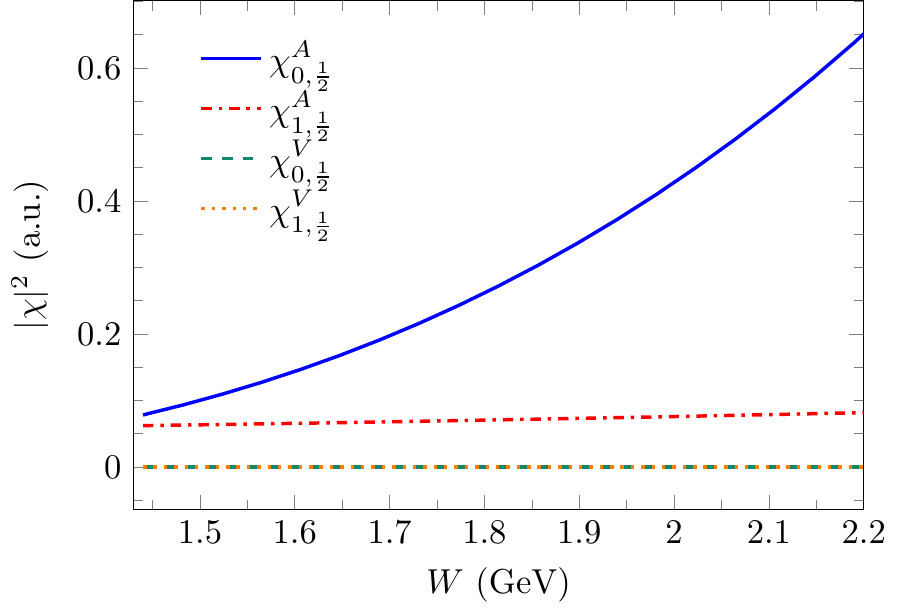} \qquad%
\includegraphics[width=0.45\textwidth]{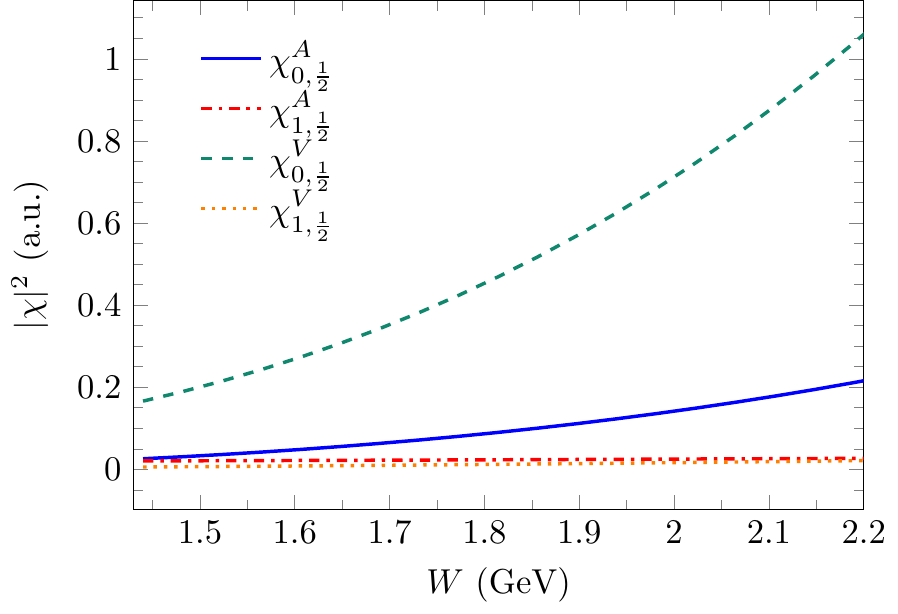} \\%
\caption{Absolute value squared of the CT contribution to $\chi^{V,A}_{r,\lambda}$, defined using  Eqs.~\eqref{eq:chi1}, \eqref{eq::chi_V_A} and  \eqref{eq::chi_Iso}, as a function of the $KN$ invariant mass
($W$) for a fixed $Q^2=0.1$ GeV$^2$. Left and right panels stand for isospin $I=0$ and $I=1$ channels, respectively. }
\label{fig::chi} 
\end{figure*} 

\begin{figure*}
    \centering
    \includegraphics[width=0.43\textwidth]{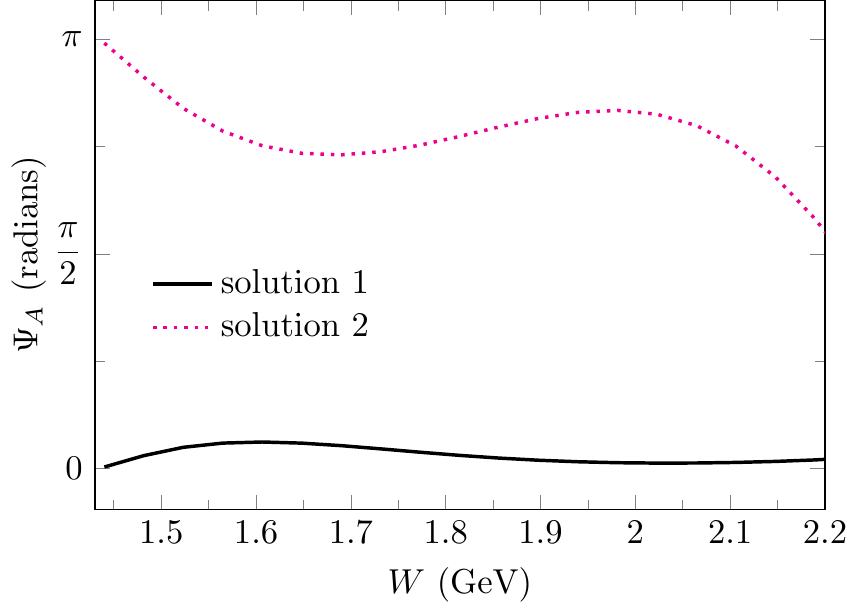} \qquad%
    \includegraphics[width=0.43\textwidth]{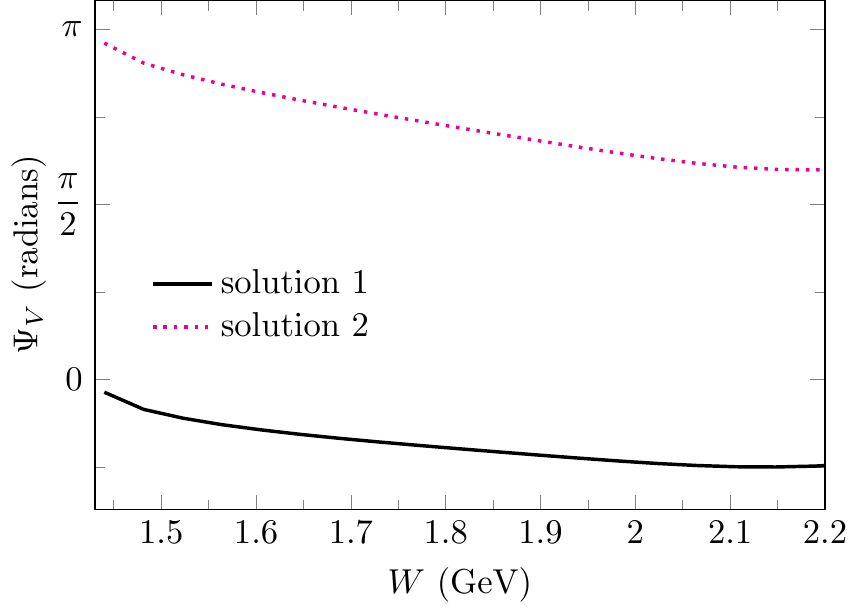}
    \caption{Olsson's phases $\Psi_{V,A}$ obtained by solving  Eqs.~\eqref{eq::chi_quad_eq1} and \eqref{eq::chi_quad_eq2} as a function of $W$ for a fixed $Q^2=0.1$ GeV$^2$.}
    \label{fig:phases}
\end{figure*}
  
\begin{figure*}
\begin{center}
\includegraphics[width=0.32\textwidth]{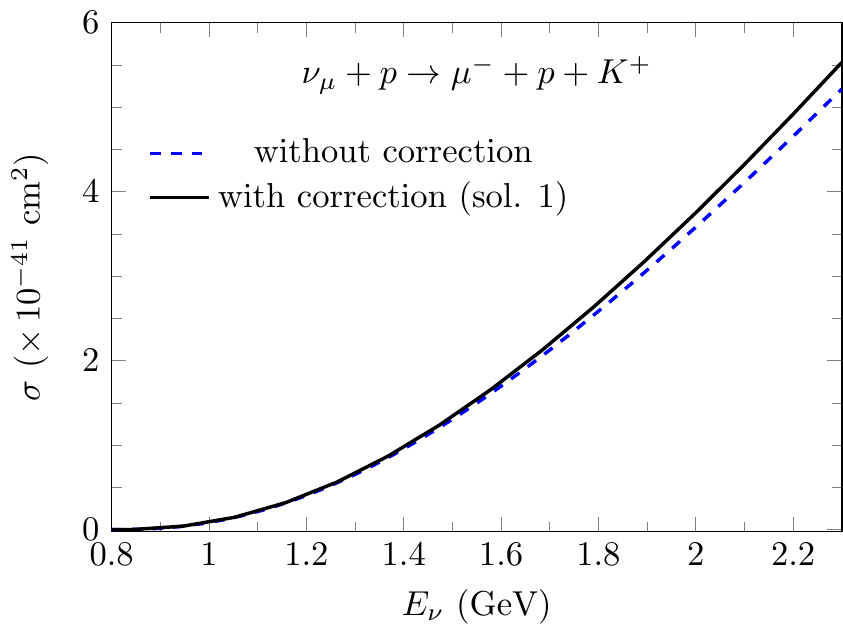} 
\includegraphics[width=0.32\textwidth]{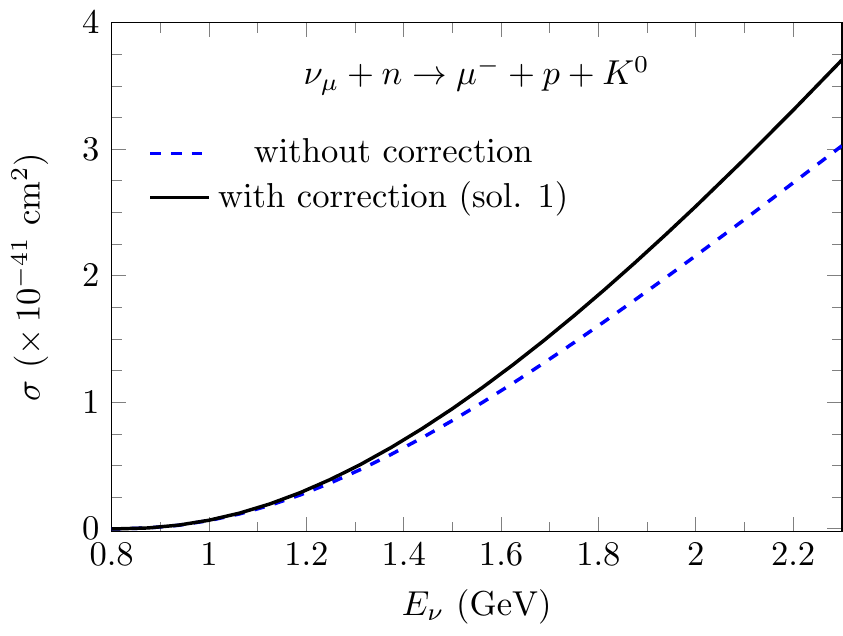} 
\includegraphics[width=0.32\textwidth]{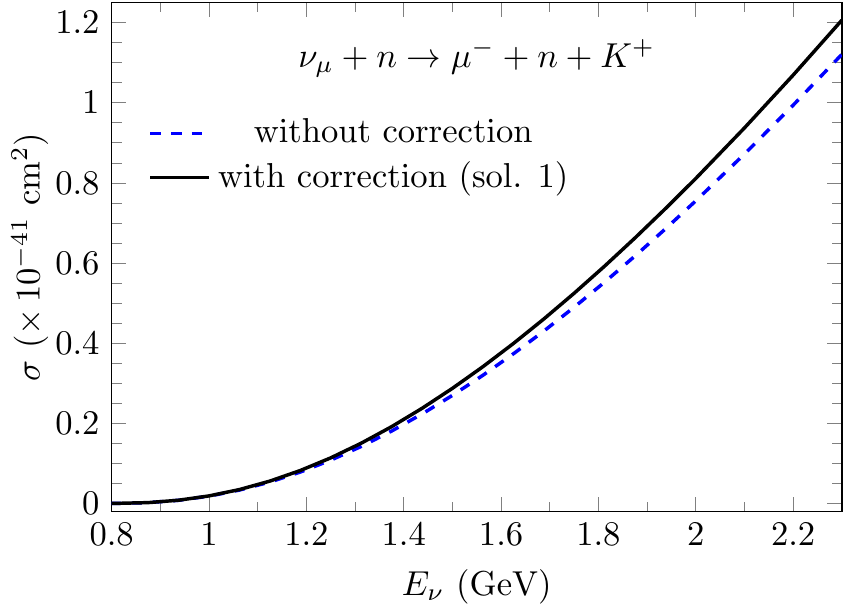}
\caption{Total cross section $\sigma(E_\nu)$ as a function of the muon-neutrino energy ($E_\nu$) for the processes of Eq.~\eqref{eq:process}. Blue dashed lines stand for the original results of Ref.~\cite{RafiAlam:2010kf}, while the predictions obtained after implementing Watson's corrections, for the chosen solution 1, are shown by the solid black lines. 
}
\label{fig:total_CS}
\end{center}
\end{figure*}  

From the analysis of Ref.~\cite{RafiAlam:2010kf} we know that contact term (CT) is the largest one for all processes in Eq.~\eqref{eq:process}. We therefore find convenient to split 
the $T$ matrix as $T=T_{CT}+T_{B}$, where $T_{CT}$ denotes the CT term,  while the rest of the diagrams of Fig.~\ref{fig:diags} are included in $T_B$. Next, we compute all the independent $\chi^{V,A\,(I=0,1)}_{r,1/2}$ with $r=0,1$ (eight in total), calculated from the CT Feynman diagram. As illustrated in Fig.~\ref{fig::chi} for a fixed $Q^2$, we identify $\chi^{A(0)}_{0,1/2}$ and $\chi^{V(1)}_{0,1/2}$ as dominant among the CT contributions, and select them to determine the Olsson's phases introduced next. 

In order to implement Watson's theorem to partially restore unitarity, we follow the prescription given by Olsson~\cite{Olsson:1974sw}. Namely, we introduce phases $\Psi_{V,A}$ in both vector and axial CT terms, such that the modified amplitude reads as
\begin{widetext}
\begin{eqnarray}
\braket{\theta, \varphi; 0, \rho | T(s) | \theta, \varphi=0; r, \lambda} 
 = \epsilon_{r\mu} T^{V\mu}_{\text{B} \lambda \rho} (\theta, \varphi) - \epsilon_{r\mu} T^{A\mu}_{\text{B} \lambda \rho} (\theta, \varphi) + \epsilon_{r\mu} T^{V\mu}_{\text{CT} \lambda \rho}(\theta, \varphi)  \, e^{i\Psi_V} - \epsilon_{r\mu} T^{A\mu}_{\text{CT} \lambda \rho}(\theta, \varphi)  \, e^{i\Psi_A} .
\label{eq::Olsson_phases}
\end{eqnarray} 
\end{widetext}
where $\epsilon_{(r,r^\prime)\mu},\, r=0,\pm1$, is the $W-$boson polarization vector. 
Thanks to  Watson's theorem these unknown phases can be determined using the available experimental information about $KN$ scattering phase shifts. We impose that
\begin{eqnarray}
    \text{Im}{\left\lbrace \chi^{V (1)}_{0,1/2}(s) \, e^{-i \delta_{S_{11}}}\right\rbrace } &=& 0 \,,  \label{eq::chi_quad_eq1} \\
    \text{Im}{\left\lbrace \chi^{A (0)}_{0,1/2}(s) \, e^{-i \delta_{S_{01}}}\right\rbrace } &=& 0 \,, \label{eq::chi_quad_eq2}
\end{eqnarray}
where the $KN$ phase shift $\delta_{L_{I,2J}}$ are taken  from the SAID database (Scattering Analyses Interactive Dialin) of the INS Data Analysis Center~\cite{SAID}. Equations~\eqref{eq::chi_quad_eq1} and \eqref{eq::chi_quad_eq2} can be used to determine Olsson's phases $\Psi_{V,A}$, which are  functions of $W$ and $Q^2$. 

\section{Results and discussion}

The $\Psi_{V,A}(W,Q^2)$ solutions of Eqs. \eqref{eq::chi_quad_eq1}, \eqref{eq::chi_quad_eq2} plugged in Eq.~\eqref{eq::Olsson_phases} correct the relative phase between the CT term and the rest of mechanisms. It should be noted, however, that these equations generally have two solutions\footnote{As discussed in Ref.~\cite{Alvarez-Ruso:2015eva} for pion production, these two solutions lead to $\chi^{V (1)}_{0,1/2}$ ($\chi^{A (0)}_{0,1/2}$) with phases $\delta_{S_{11}(S_{01})}$ and $\delta_{S_{11}(S_{01})} + \pi$ ($KN$ phase shifts are defined up to a summand of $\pi$).} denoted here as solutions 1 and 2. The $W$ dependence of these phases  is shown in Fig. \ref{fig:phases} for the same fixed $Q^2$ used in Fig. \ref{fig::chi}. The plots show the general tendency for solution 1 (2) to be small (large) phases in the range of $KN$ invariant masses under consideration. The four combinations of Olsson's phases $\Psi_{V,A}(W,Q^2)$ that can be assembled with these two solutions lead to different values for observable quantities. In Ref.~\cite{Alvarez-Ruso:2015eva}, where a similar approach was undertaken for weak pion production, the preference for small Olsson's phases was clearly validated by pion photoproduction data (see Fig. 2 of that paper). In the present case, there are no equivalent electromagnetic single kaon production data that could serve for validation purposes. However, as illustrated in Fig. \ref{fig:phases}, at low $W$ and $Q^2$, i.e. close to threshold, $\Psi_{V,A} \sim \pi$ for solution 2. Such a behavior implies a relative sign between $T_{CT}$ and $T_B$ which is inconsistent with the predictions of chiral symmetry encoded in the leading-order Lagrangian. We thus rely on this observation to discard solution 2 in our predictions.   

The integrated cross sections obtained with solution 1 are shown in Fig.~\ref{fig:total_CS},  together with the reference calculation of Ref.~\cite{RafiAlam:2010kf}, which did not include the Olsson's phases. One immediately notices that the partial unitarization  causes a small variation in the cross section.  The largest change, observed in $\nu_\mu n \rightarrow \mu^- p K^0$, amounts to about an 18\%  increase with respect to the reference predictions of Ref.~\cite{RafiAlam:2010kf}  at $E_\nu=2$ GeV. This small effect is plausibly a consequence of the weakness (for strong forces) of the $KN$ interactions. One can therefore expect that, in the energy region in which the present model is applicable, the size of unitarity corrections is within the model uncertainties (effectively accounted by the 10~\% uncertainty assumed for the dipole mass) at least for the total cross section.
Future data for weak single kaon production at low energies obtained, for example with the Short Baseline Near Detector (SBND)~\cite{Antonello:2015lea} at Fermilab, that will collect data with high statistics, or in a future neutrino experiment on hydrogen and/or deuterium could be compared to our predictions, shedding light on this interesting process.

\begin{figure*}
    \centering
    \includegraphics[width=0.87\textwidth]{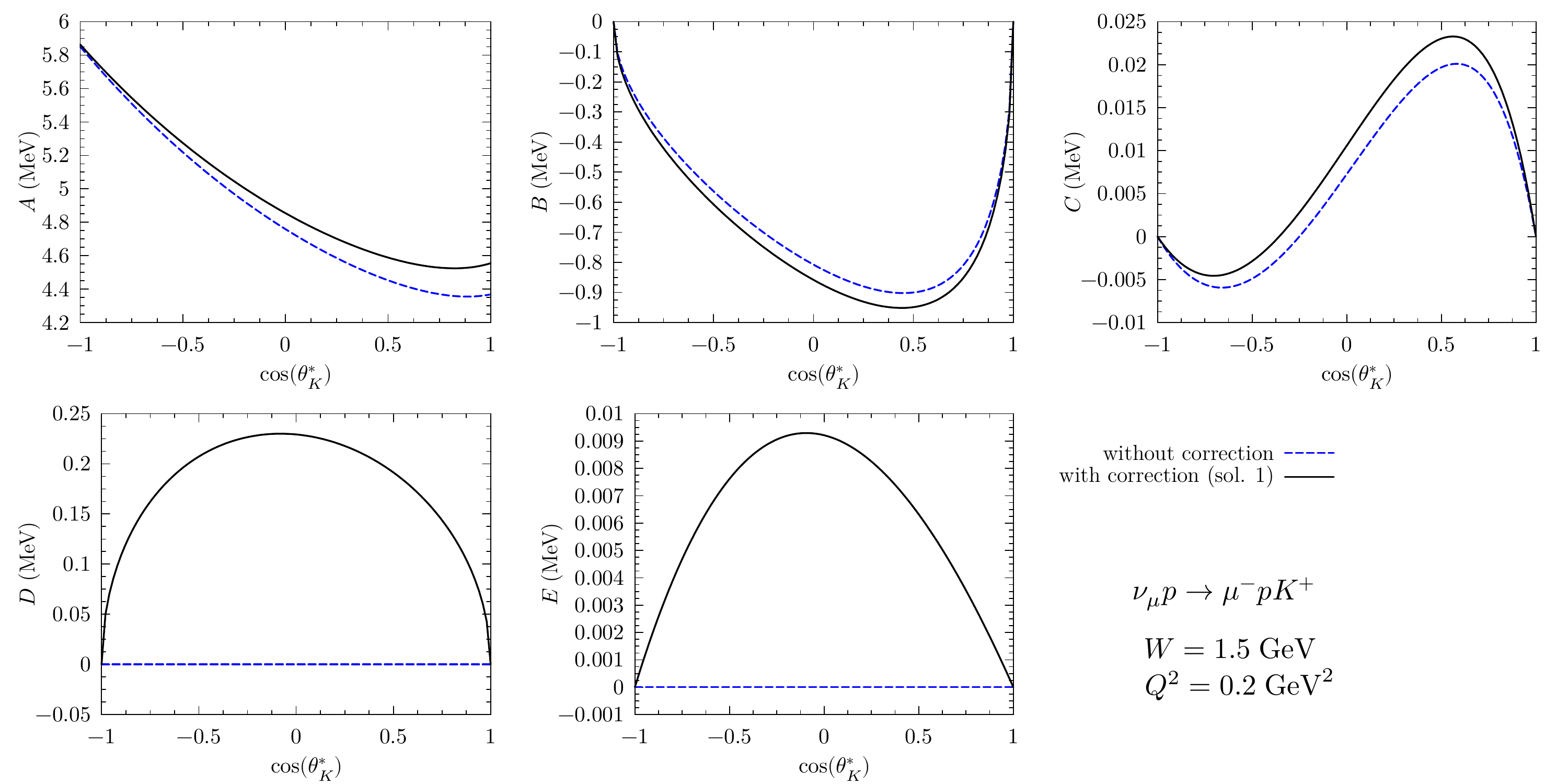}
    \includegraphics[width=0.87\textwidth]{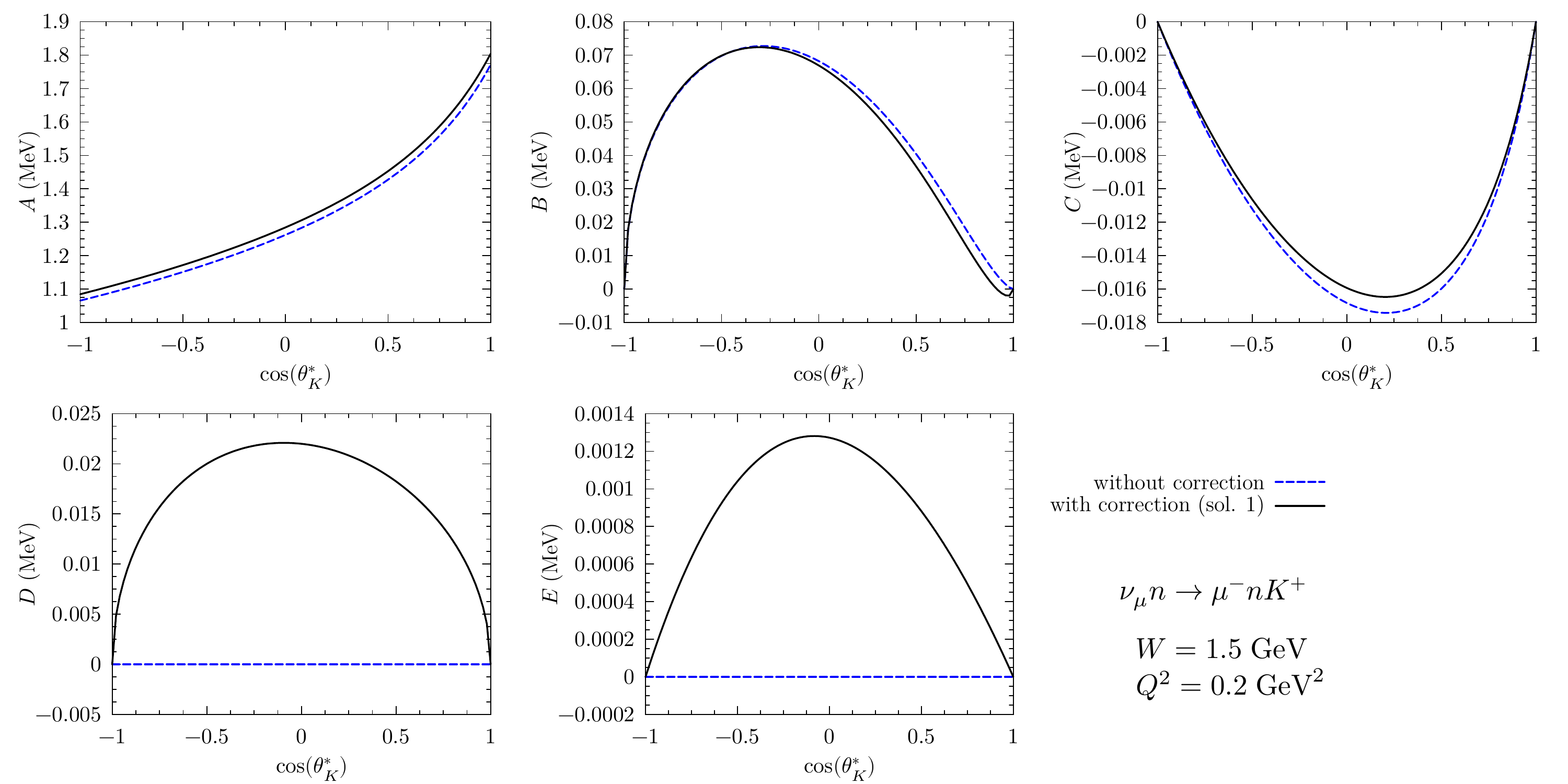}
    \includegraphics[width=0.87\textwidth]{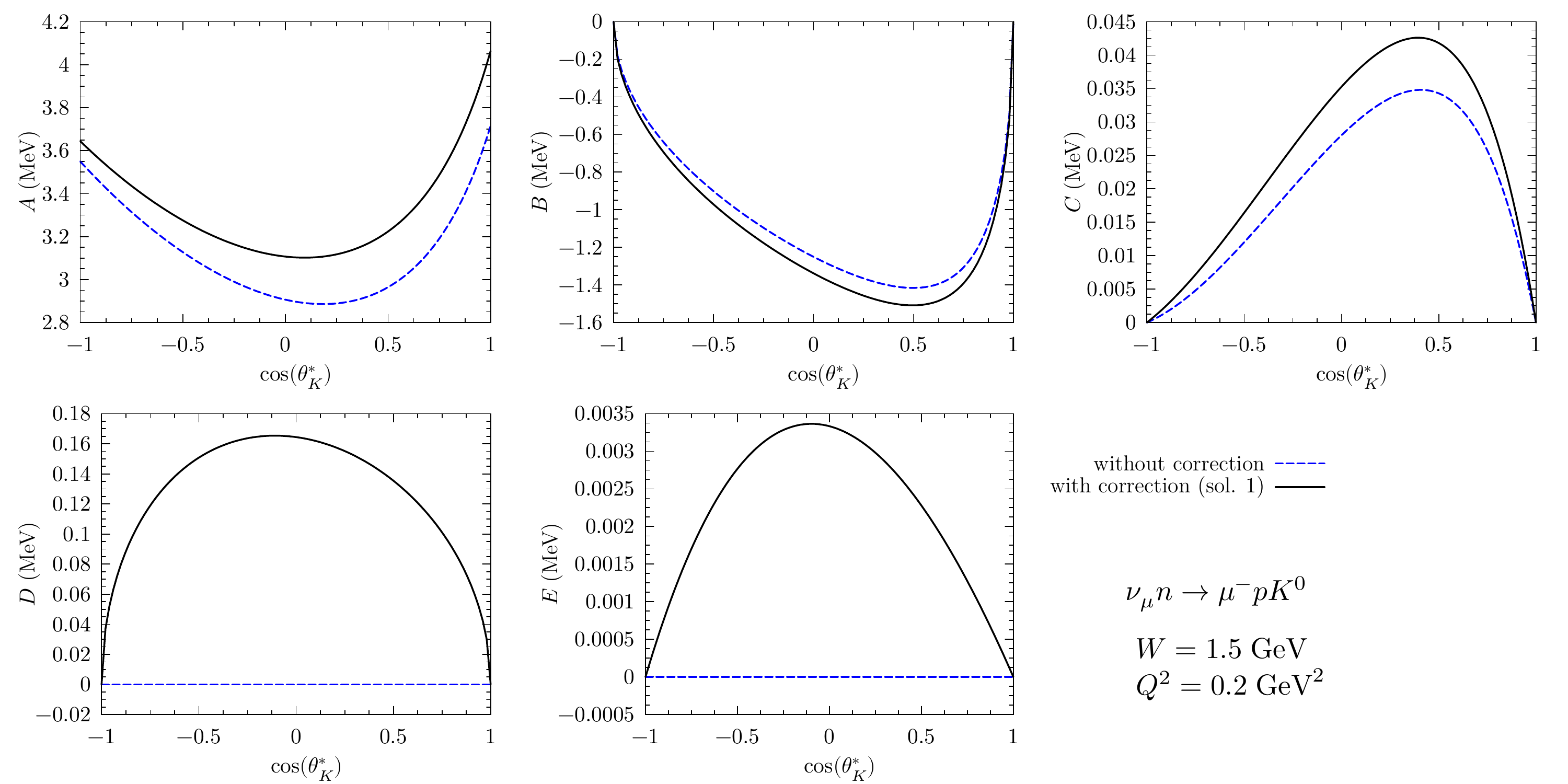}
    \caption{$A,B,C,D,E$ structure functions for $\nu_\mu + N \rightarrow \mu^- + N' + K$ as a function of the cosine of the polar kaon angle in the HCM frame ($\theta^*_K$) for fixed $E_\nu = 2$ GeV, $W=1.5$ GeV and $Q^2=0.2$ GeV$^2$. }
    \label{fig:st_fun}
\end{figure*}

In order to perform a more detailed analysis of the impact of unitarity corrections we rely on the following representation of the differential cross section, Eq.~\eqref{eq:diff_phsp}, 
\begin{equation}
\begin{aligned}
    \frac{d^4 \sigma}{ dW \, d Q^2 d\Omega^*_K} =& \frac{G_F^2 W}{4 \pi M_N |\Vec{k}|^2} \left( A + B \cos \phi^*_K + C \cos 2\phi^*_K \right.
    \\
    & \left.  + D \sin \phi^*_K + E \sin 2\phi^*_K \right) \,,
\end{aligned}
\end{equation}
where the dependence on the HCM kaon azimuthal angle has been singled out~\cite{Sobczyk:2018ghy,Hernandez:2007qq,Hernandez:2006yg}. The incoming neutrino momentum $\vec k$ is in the Laboratory frame while kaon angles (carrying the `*' superscript) are in the HCM frame. The structure functions $A-E$ are real and depend on the scalars $Q^2$, $p\cdot q$, $p_K \cdot q$ and $p_K \cdot p$. We have obtained these structure functions  for weak kaon production for the first time. They are displayed in Fig.~\ref{fig:st_fun} as a function of $\cos{\theta^*_K}$ for fixed $E_\nu$, $W$ and $Q^2$. Results obtained with solution 1 are close to the uncorrected ones as expected.
Remarkably, the $D$ and $E$ structure functions, responsible for parity violation in kaon production (and weak meson production in general~\cite{Hernandez:2006yg}), which are zero in the tree-level model with real amplitudes, acquire nonzero although small values due to unitarization.     

\section{Conclusion}

We have improved the theoretical description of single kaon production in neutrino-nucleon collisions below the $KKY$ threshold by partially accounting for unitarity. For this purpose we have introduced Olsson's phases for the contact term of the amplitude in its largest vector and axial multipoles. These phases take the values required to fulfill Watson's theorem. In the absence of experimental data, we have relied on chiral symmetry to discard some of the found mathematical solutions. 
The remaining solution leads to small corrections in the cross section, as expected because of the absence of baryon
resonances. These corrections are actually within the uncertainties of the model. This would validate the reference tree-level model, built upon the leading-order chiral Lagrangian, in the kinematic region under consideration. Finally, we have investigated the behavior of the structure functions that characterize the cross-section dependence on the kaon azimuthal angle. The impact of unitarization is visible in the fact that the parity-violating  structure functions depart from zero. 

\section*{Acknowledgements}

We thank E. Hern\'andez for useful feedback. MRA is thankful to IFIC, Valencia for the hospitality during his stay. This research has been partially supported by Spanish Ministerio de Ciencia e Innovaci\'on and the European Regional Development Fund (ERDF) under contract FIS2017-84038-C2-1-P, the EU STRONG-2020 project under the program H2020-INFRAIA-2018-1, grant agreement no. 824093, by Generalitat Valenciana under contract PROMETEO/2020/023, and by the Deutsche Forschungsgemeinschaft (DFG) through the Collaborative
 Research Center [The Low-Energy Frontier of the Standard Model (SFB 1044)] and through the Cluster of Excellence ``Precision Physics, Fundamental Interactions, and Structure of Matter" (PRISMA$^+$ EXC 2118/1) funded by the DFG within the German Excellence Strategy (Project ID 39083149).

\appendix
\section*{Appendices}
\addcontentsline{toc}{section}{Appendices}
\renewcommand{\thesubsection}{\Alph{subsection}} 
\setcounter{equation}{0}
\renewcommand{\theequation}{A\arabic{equation}}

\subsection{Basis transformations}\label{app:states}

 The states with well defined total angular momentum and the two-particle helicity states are related by the transformation relation:
 \begin{equation}
 \begin{aligned}
&\ket{J,M_J; \lambda_1,\lambda_2} 
\\
&= \sqrt{\frac{2J+1}{4\pi}} \int d\Omega \, \mathcal{D}_{M_J \lambda}^{(J)*}\left(\phi_K, \theta_K, -\phi_K\right) \ket{\theta_K,\phi_K; \lambda_1,\lambda_2}
\end{aligned}
\label{eq::relation_between_states_noCJ}
 \end{equation}
with $\lambda=\lambda_1 - \lambda_2$. $\mathcal{D}^{(J)}_{M_J \lambda} \left( \alpha,\beta,\gamma \right)$ is the  Wigner rotation matrix.

In the $L$-$S$ scheme, where we use the basis $\ket{J,M_J; L,S}$ with $L$ the orbital angular momentum and $S$ the total spin of the two particles, the following relations hold
\begin{equation}\label{eq:trans1}
\begin{aligned}
\ket{J,M_J;\lambda_1, \lambda_2} = \sum_{L,S} \sqrt{\frac{2L+1}{2J+1}} \left(L,S,J | 0,\lambda,\lambda\right)\\
 \times\left(j_1,j_2,S| \lambda_1,-\lambda_2,\lambda\right) \ket{J,M_J;L,S} \,,
\\
\ket{J,M_J;L,S} = \sum_{\lambda_1, \lambda_2} \sqrt{\frac{2L+1}{2J+1}} \left(L,S,J | 0,\lambda,\lambda\right) \\
 \times\left(j_1,j_2,S| \lambda_1,-\lambda_2,\lambda\right) \ket{J,M_J;\lambda_1, \lambda_2}\,,
\end{aligned}
\end{equation}
where $j_i$ is the total angular momentum of each particle and $\left(j_1,j_2,J| m_1,m_2,M\right) $ are Clebsch-Gordan coefficients.

\subsection{Properties of $\chi^{V,A}_{r,\lambda}$ functions under helicity inversion}\label{app::parity}

In terms of two-particle helicity states with well defined angular momentum $J$ ($=1/2$ in our case)
\begin{equation}
\chi^{V,A}_{r,\lambda} = \sum_\rho \bra{1/2,M;0,\rho} T^{V,A} \ket{1/2,M;r,\lambda} \,.
\end{equation}
Under parity inversion, these states are transformed as (Eq. (5.28) of Ref. \cite{Martin:102663}) 
\begin{equation}
    P \ket{J,M;\mu_1,\mu_2} = \eta_1 \eta_2 (-1)^{J-s_1 -s_2} \ket{J,M;-\mu_1,-\mu_2}
    \nonumber
\end{equation}
in terms of the two particles' intrinsic parities $\eta_{1,2}$ and spins $s_{1,2}$. Therefore
\begin{eqnarray}
    P \ket{1/2,M;r,\lambda} &=& \eta_N \eta_W (-1)^{1/2 - 1/2 - 1} \ket{1/2,M;-r,-\lambda}\, , \nonumber \\
    P \ket{1/2,M;0,\rho} &=& \eta_N \eta_K (-1)^{1/2 - 1/2 - 0} \ket{1/2,M;-r,-\lambda}\, . \nonumber 
\end{eqnarray}
Consequently 
\begin{equation}
 \chi^{V,A}_{-r,-\lambda} = - \sum_\rho \bra{1/2,M;0,\rho} P^{-1} T^{V,A} P \ket{1/2,M;r,\lambda} \nonumber \,,
\end{equation}
where we have taken into account that these matrix elements do not depend on $M$ because $T$ is a scalar under rotations. Once  $P^{-1} T^{V,A} P = \pm T^{V,A}$
\begin{equation}
 \chi^{V,A}_{-r,-\lambda} = \mp  \chi^{V,A}_{r,\lambda} \,,  
\end{equation}
from where Eq.~\eqref{eq::chi_V_A} immediately follows.     

\newpage
 \bibliography{biblio}

\begin{thebibliography}{27}%
\makeatletter
\providecommand \@ifxundefined [1]{%
 \@ifx{#1\undefined}
}%
\providecommand \@ifnum [1]{%
 \ifnum #1\expandafter \@firstoftwo
 \else \expandafter \@secondoftwo
 \fi
}%
\providecommand \@ifx [1]{%
 \ifx #1\expandafter \@firstoftwo
 \else \expandafter \@secondoftwo
 \fi
}%
\providecommand \natexlab [1]{#1}%
\providecommand \enquote  [1]{``#1''}%
\providecommand \bibnamefont  [1]{#1}%
\providecommand \bibfnamefont [1]{#1}%
\providecommand \citenamefont [1]{#1}%
\providecommand \href@noop [0]{\@secondoftwo}%
\providecommand \href [0]{\begingroup \@sanitize@url \@href}%
\providecommand \@href[1]{\@@startlink{#1}\@@href}%
\providecommand \@@href[1]{\endgroup#1\@@endlink}%
\providecommand \@sanitize@url [0]{\catcode `\\12\catcode `\$12\catcode
  `\&12\catcode `\#12\catcode `\^12\catcode `\_12\catcode `\%12\relax}%
\providecommand \@@startlink[1]{}%
\providecommand \@@endlink[0]{}%
\providecommand \url  [0]{\begingroup\@sanitize@url \@url }%
\providecommand \@url [1]{\endgroup\@href {#1}{\urlprefix }}%
\providecommand \urlprefix  [0]{URL }%
\providecommand \Eprint [0]{\href }%
\providecommand \doibase [0]{http://dx.doi.org/}%
\providecommand \selectlanguage [0]{\@gobble}%
\providecommand \bibinfo  [0]{\@secondoftwo}%
\providecommand \bibfield  [0]{\@secondoftwo}%
\providecommand \translation [1]{[#1]}%
\providecommand \BibitemOpen [0]{}%
\providecommand \bibitemStop [0]{}%
\providecommand \bibitemNoStop [0]{.\EOS\space}%
\providecommand \EOS [0]{\spacefactor3000\relax}%
\providecommand \BibitemShut  [1]{\csname bibitem#1\endcsname}%
\let\auto@bib@innerbib\@empty
\bibitem [{\citenamefont {Rafi~Alam}\ \emph {et~al.}(2010)\citenamefont
  {Rafi~Alam}, \citenamefont {Ruiz~Simo}, \citenamefont {Sajjad~Athar},\ and\
  \citenamefont {Vicente~Vacas}}]{RafiAlam:2010kf}%
  \BibitemOpen
  \bibfield  {author} {\bibinfo {author} {\bibfnamefont {M.}~\bibnamefont
  {Rafi~Alam}}, \bibinfo {author} {\bibfnamefont {I.}~\bibnamefont
  {Ruiz~Simo}}, \bibinfo {author} {\bibfnamefont {M.}~\bibnamefont
  {Sajjad~Athar}}, \ and\ \bibinfo {author} {\bibfnamefont {M.~J.}\
  \bibnamefont {Vicente~Vacas}},\ }\href {\doibase 10.1103/PhysRevD.82.033001}
  {\bibfield  {journal} {\bibinfo  {journal} {Phys. Rev.}\ }\textbf {\bibinfo
  {volume} {D82}},\ \bibinfo {pages} {033001} (\bibinfo {year} {2010})},\
  \Eprint {http://arxiv.org/abs/1004.5484} {arXiv:1004.5484 [hep-ph]}
  \BibitemShut {NoStop}%
\bibitem [{\citenamefont {Mahn}\ \emph {et~al.}(2018)\citenamefont {Mahn},
  \citenamefont {Marshall},\ and\ \citenamefont {Wilkinson}}]{Mahn:2018mai}%
  \BibitemOpen
  \bibfield  {author} {\bibinfo {author} {\bibfnamefont {K.}~\bibnamefont
  {Mahn}}, \bibinfo {author} {\bibfnamefont {C.}~\bibnamefont {Marshall}}, \
  and\ \bibinfo {author} {\bibfnamefont {C.}~\bibnamefont {Wilkinson}},\ }\href
  {\doibase 10.1146/annurev-nucl-101917-020930} {\bibfield  {journal} {\bibinfo
   {journal} {Ann. Rev. Nucl. Part. Sci.}\ }\textbf {\bibinfo {volume} {68}},\
  \bibinfo {pages} {105} (\bibinfo {year} {2018})},\ \Eprint
  {http://arxiv.org/abs/1803.08848} {arXiv:1803.08848 [hep-ex]} \BibitemShut
  {NoStop}%
\bibitem [{\citenamefont {Alvarez-Ruso}\ \emph {et~al.}(2018)\citenamefont
  {Alvarez-Ruso} \emph {et~al.}}]{Alvarez-Ruso:2017oui}%
  \BibitemOpen
  \bibfield  {author} {\bibinfo {author} {\bibfnamefont {L.}~\bibnamefont
  {Alvarez-Ruso}} \emph {et~al.},\ }\href {\doibase 10.1016/j.ppnp.2018.01.006}
  {\bibfield  {journal} {\bibinfo  {journal} {Prog. Part. Nucl. Phys.}\
  }\textbf {\bibinfo {volume} {100}},\ \bibinfo {pages} {1} (\bibinfo {year}
  {2018})},\ \Eprint {http://arxiv.org/abs/1706.03621} {arXiv:1706.03621
  [hep-ph]} \BibitemShut {NoStop}%
\bibitem [{\citenamefont {Katori}\ and\ \citenamefont
  {Martini}(2018)}]{Katori:2016yel}%
  \BibitemOpen
  \bibfield  {author} {\bibinfo {author} {\bibfnamefont {T.}~\bibnamefont
  {Katori}}\ and\ \bibinfo {author} {\bibfnamefont {M.}~\bibnamefont
  {Martini}},\ }\href {\doibase 10.1088/1361-6471/aa8bf7} {\bibfield  {journal}
  {\bibinfo  {journal} {J. Phys.}\ }\textbf {\bibinfo {volume} {G45}},\
  \bibinfo {pages} {013001} (\bibinfo {year} {2018})},\ \Eprint
  {http://arxiv.org/abs/1611.07770} {arXiv:1611.07770 [hep-ph]} \BibitemShut
  {NoStop}%
\bibitem [{\citenamefont {Alvarez-Ruso}\ \emph {et~al.}(2014)\citenamefont
  {Alvarez-Ruso}, \citenamefont {Hayato},\ and\ \citenamefont
  {Nieves}}]{Alvarez-Ruso:2014bla}%
  \BibitemOpen
  \bibfield  {author} {\bibinfo {author} {\bibfnamefont {L.}~\bibnamefont
  {Alvarez-Ruso}}, \bibinfo {author} {\bibfnamefont {Y.}~\bibnamefont
  {Hayato}}, \ and\ \bibinfo {author} {\bibfnamefont {J.}~\bibnamefont
  {Nieves}},\ }\href {\doibase 10.1088/1367-2630/16/7/075015} {\bibfield
  {journal} {\bibinfo  {journal} {New J. Phys.}\ }\textbf {\bibinfo {volume}
  {16}},\ \bibinfo {pages} {075015} (\bibinfo {year} {2014})},\ \Eprint
  {http://arxiv.org/abs/1403.2673} {arXiv:1403.2673 [hep-ph]} \BibitemShut
  {NoStop}%
\bibitem [{\citenamefont {Formaggio}\ and\ \citenamefont
  {Zeller}(2012)}]{Formaggio:2013kya}%
  \BibitemOpen
  \bibfield  {author} {\bibinfo {author} {\bibfnamefont {J.~A.}\ \bibnamefont
  {Formaggio}}\ and\ \bibinfo {author} {\bibfnamefont {G.~P.}\ \bibnamefont
  {Zeller}},\ }\href {\doibase 10.1103/RevModPhys.84.1307} {\bibfield
  {journal} {\bibinfo  {journal} {Rev. Mod. Phys.}\ }\textbf {\bibinfo {volume}
  {84}},\ \bibinfo {pages} {1307} (\bibinfo {year} {2012})},\ \Eprint
  {http://arxiv.org/abs/1305.7513} {arXiv:1305.7513 [hep-ex]} \BibitemShut
  {NoStop}%
\bibitem [{\citenamefont {Marshall}\ \emph {et~al.}(2016)\citenamefont
  {Marshall} \emph {et~al.}}]{Marshall:2016rrn}%
  \BibitemOpen
  \bibfield  {author} {\bibinfo {author} {\bibfnamefont {C.~M.}\ \bibnamefont
  {Marshall}} \emph {et~al.} (\bibinfo {collaboration} {MINERvA}),\ }\href
  {\doibase 10.1103/PhysRevD.94.012002} {\bibfield  {journal} {\bibinfo
  {journal} {Phys. Rev.}\ }\textbf {\bibinfo {volume} {D94}},\ \bibinfo {pages}
  {012002} (\bibinfo {year} {2016})},\ \Eprint
  {http://arxiv.org/abs/1604.03920} {arXiv:1604.03920 [hep-ex]} \BibitemShut
  {NoStop}%
\bibitem [{\citenamefont {Lalakulich}\ \emph {et~al.}(2012)\citenamefont
  {Lalakulich}, \citenamefont {Gallmeister},\ and\ \citenamefont
  {Mosel}}]{Lalakulich:2012gm}%
  \BibitemOpen
  \bibfield  {author} {\bibinfo {author} {\bibfnamefont {O.}~\bibnamefont
  {Lalakulich}}, \bibinfo {author} {\bibfnamefont {K.}~\bibnamefont
  {Gallmeister}}, \ and\ \bibinfo {author} {\bibfnamefont {U.}~\bibnamefont
  {Mosel}},\ }\href {\doibase 10.1103/PhysRevC.86.014607} {\bibfield  {journal}
  {\bibinfo  {journal} {Phys. Rev.}\ }\textbf {\bibinfo {volume} {C86}},\
  \bibinfo {pages} {014607} (\bibinfo {year} {2012})},\ \Eprint
  {http://arxiv.org/abs/1205.1061} {arXiv:1205.1061 [nucl-th]} \BibitemShut
  {NoStop}%
\bibitem [{\citenamefont {Shrock}(1975)}]{Shrock:1975an}%
  \BibitemOpen
  \bibfield  {author} {\bibinfo {author} {\bibfnamefont {R.~E.}\ \bibnamefont
  {Shrock}},\ }\href {\doibase 10.1103/PhysRevD.12.2049} {\bibfield  {journal}
  {\bibinfo  {journal} {Phys. Rev.}\ }\textbf {\bibinfo {volume} {D12}},\
  \bibinfo {pages} {2049} (\bibinfo {year} {1975})}\BibitemShut {NoStop}%
\bibitem [{\citenamefont {Mecklenburg}(1978)}]{Mecklenburg:1976pk}%
  \BibitemOpen
  \bibfield  {author} {\bibinfo {author} {\bibfnamefont {W.}~\bibnamefont
  {Mecklenburg}},\ }\href@noop {} {\bibfield  {journal} {\bibinfo  {journal}
  {Acta Phys. Austriaca}\ }\textbf {\bibinfo {volume} {48}},\ \bibinfo {pages}
  {293} (\bibinfo {year} {1978})}\BibitemShut {NoStop}%
\bibitem [{\citenamefont {Amer}(1978)}]{Amer:1977fy}%
  \BibitemOpen
  \bibfield  {author} {\bibinfo {author} {\bibfnamefont {A.~A.}\ \bibnamefont
  {Amer}},\ }\href {\doibase 10.1103/PhysRevD.18.2290} {\bibfield  {journal}
  {\bibinfo  {journal} {Phys. Rev.}\ }\textbf {\bibinfo {volume} {D18}},\
  \bibinfo {pages} {2290} (\bibinfo {year} {1978})}\BibitemShut {NoStop}%
\bibitem [{\citenamefont {Dewan}(1981)}]{Dewan:1981ab}%
  \BibitemOpen
  \bibfield  {author} {\bibinfo {author} {\bibfnamefont {H.~K.}\ \bibnamefont
  {Dewan}},\ }\href {\doibase 10.1103/PhysRevD.24.2369} {\bibfield  {journal}
  {\bibinfo  {journal} {Phys. Rev.}\ }\textbf {\bibinfo {volume} {D24}},\
  \bibinfo {pages} {2369} (\bibinfo {year} {1981})}\BibitemShut {NoStop}%
\bibitem [{\citenamefont {Adera}\ \emph {et~al.}(2010)\citenamefont {Adera},
  \citenamefont {Van Der~Ventel}, \citenamefont {van Niekerk},\ and\
  \citenamefont {Mart}}]{Adera:2010zz}%
  \BibitemOpen
  \bibfield  {author} {\bibinfo {author} {\bibfnamefont {G.~B.}\ \bibnamefont
  {Adera}}, \bibinfo {author} {\bibfnamefont {B.~I.~S.}\ \bibnamefont {Van
  Der~Ventel}}, \bibinfo {author} {\bibfnamefont {D.~D.}\ \bibnamefont {van
  Niekerk}}, \ and\ \bibinfo {author} {\bibfnamefont {T.}~\bibnamefont
  {Mart}},\ }\href {\doibase 10.1103/PhysRevC.82.025501} {\bibfield  {journal}
  {\bibinfo  {journal} {Phys. Rev.}\ }\textbf {\bibinfo {volume} {C82}},\
  \bibinfo {pages} {025501} (\bibinfo {year} {2010})},\ \Eprint
  {http://arxiv.org/abs/1112.5748} {arXiv:1112.5748 [nucl-th]} \BibitemShut
  {NoStop}%
\bibitem [{\citenamefont {Nakamura}\ \emph {et~al.}(2015)\citenamefont
  {Nakamura}, \citenamefont {Kamano},\ and\ \citenamefont
  {Sato}}]{Nakamura:2015rta}%
  \BibitemOpen
  \bibfield  {author} {\bibinfo {author} {\bibfnamefont {S.~X.}\ \bibnamefont
  {Nakamura}}, \bibinfo {author} {\bibfnamefont {H.}~\bibnamefont {Kamano}}, \
  and\ \bibinfo {author} {\bibfnamefont {T.}~\bibnamefont {Sato}},\ }\href
  {\doibase 10.1103/PhysRevD.92.074024} {\bibfield  {journal} {\bibinfo
  {journal} {Phys. Rev.}\ }\textbf {\bibinfo {volume} {D92}},\ \bibinfo {pages}
  {074024} (\bibinfo {year} {2015})},\ \Eprint
  {http://arxiv.org/abs/1506.03403} {arXiv:1506.03403 [hep-ph]} \BibitemShut
  {NoStop}%
\bibitem [{\citenamefont {Alam}\ \emph {et~al.}(2012)\citenamefont {Alam},
  \citenamefont {Simo}, \citenamefont {Athar},\ and\ \citenamefont
  {Vicente~Vacas}}]{Alam:2011xq}%
  \BibitemOpen
  \bibfield  {author} {\bibinfo {author} {\bibfnamefont {M.~R.}\ \bibnamefont
  {Alam}}, \bibinfo {author} {\bibfnamefont {I.~R.}\ \bibnamefont {Simo}},
  \bibinfo {author} {\bibfnamefont {M.~S.}\ \bibnamefont {Athar}}, \ and\
  \bibinfo {author} {\bibfnamefont {M.~J.}\ \bibnamefont {Vicente~Vacas}},\
  }\href {\doibase 10.1103/PhysRevD.85.013014} {\bibfield  {journal} {\bibinfo
  {journal} {Phys. Rev.}\ }\textbf {\bibinfo {volume} {D85}},\ \bibinfo {pages}
  {013014} (\bibinfo {year} {2012})},\ \Eprint {http://arxiv.org/abs/1111.0863}
  {arXiv:1111.0863 [hep-ph]} \BibitemShut {NoStop}%
\bibitem [{\citenamefont {Ren}\ \emph {et~al.}(2015)\citenamefont {Ren},
  \citenamefont {Oset}, \citenamefont {Alvarez-Ruso},\ and\ \citenamefont
  {Vicente~Vacas}}]{Ren:2015bsa}%
  \BibitemOpen
  \bibfield  {author} {\bibinfo {author} {\bibfnamefont {X.-L.}\ \bibnamefont
  {Ren}}, \bibinfo {author} {\bibfnamefont {E.}~\bibnamefont {Oset}}, \bibinfo
  {author} {\bibfnamefont {L.}~\bibnamefont {Alvarez-Ruso}}, \ and\ \bibinfo
  {author} {\bibfnamefont {M.~J.}\ \bibnamefont {Vicente~Vacas}},\ }\href
  {\doibase 10.1103/PhysRevC.91.045201} {\bibfield  {journal} {\bibinfo
  {journal} {Phys. Rev.}\ }\textbf {\bibinfo {volume} {C91}},\ \bibinfo {pages}
  {045201} (\bibinfo {year} {2015})},\ \Eprint
  {http://arxiv.org/abs/1501.04073} {arXiv:1501.04073 [hep-ph]} \BibitemShut
  {NoStop}%
\bibitem [{\citenamefont {Watson}(1952)}]{Watson:1952ji}%
  \BibitemOpen
  \bibfield  {author} {\bibinfo {author} {\bibfnamefont {K.~M.}\ \bibnamefont
  {Watson}},\ }\href {\doibase 10.1103/PhysRev.88.1163} {\bibfield  {journal}
  {\bibinfo  {journal} {Phys. Rev.}\ }\textbf {\bibinfo {volume} {88}},\
  \bibinfo {pages} {1163} (\bibinfo {year} {1952})},\ \bibinfo {note} {[Riv.
  Nuovo Cim.31,1(2008)]}\BibitemShut {NoStop}%
\bibitem [{\citenamefont {Olsson}(1974)}]{Olsson:1974sw}%
  \BibitemOpen
  \bibfield  {author} {\bibinfo {author} {\bibfnamefont {M.~G.}\ \bibnamefont
  {Olsson}},\ }\href {\doibase 10.1016/0550-3213(74)90115-1} {\bibfield
  {journal} {\bibinfo  {journal} {Nucl. Phys.}\ }\textbf {\bibinfo {volume}
  {B78}},\ \bibinfo {pages} {55} (\bibinfo {year} {1974})}\BibitemShut
  {NoStop}%
\bibitem [{\citenamefont {Alvarez-Ruso}\ \emph {et~al.}(2016)\citenamefont
  {Alvarez-Ruso}, \citenamefont {Hernández}, \citenamefont {Nieves},\ and\
  \citenamefont {Vicente~Vacas}}]{Alvarez-Ruso:2015eva}%
  \BibitemOpen
  \bibfield  {author} {\bibinfo {author} {\bibfnamefont {L.}~\bibnamefont
  {Alvarez-Ruso}}, \bibinfo {author} {\bibfnamefont {E.}~\bibnamefont
  {Hernández}}, \bibinfo {author} {\bibfnamefont {J.}~\bibnamefont {Nieves}},
  \ and\ \bibinfo {author} {\bibfnamefont {M.~J.}\ \bibnamefont
  {Vicente~Vacas}},\ }\href {\doibase 10.1103/PhysRevD.93.014016} {\bibfield
  {journal} {\bibinfo  {journal} {Phys. Rev.}\ }\textbf {\bibinfo {volume}
  {D93}},\ \bibinfo {pages} {014016} (\bibinfo {year} {2016})},\ \Eprint
  {http://arxiv.org/abs/1510.06266} {arXiv:1510.06266 [hep-ph]} \BibitemShut
  {NoStop}%
\bibitem [{\citenamefont {Hern\'andez}\ and\ \citenamefont
  {Nieves}(2017)}]{Hernandez:2016yfb}%
  \BibitemOpen
  \bibfield  {author} {\bibinfo {author} {\bibfnamefont {E.}~\bibnamefont
  {Hern\'andez}}\ and\ \bibinfo {author} {\bibfnamefont {J.}~\bibnamefont
  {Nieves}},\ }\href {\doibase 10.1103/PhysRevD.95.053007} {\bibfield
  {journal} {\bibinfo  {journal} {Phys. Rev. D}\ }\textbf {\bibinfo {volume}
  {95}},\ \bibinfo {pages} {053007} (\bibinfo {year} {2017})},\ \Eprint
  {http://arxiv.org/abs/1612.02343} {arXiv:1612.02343 [hep-ph]} \BibitemShut
  {NoStop}%
\bibitem [{\citenamefont {Hernandez}\ \emph
  {et~al.}(2007{\natexlab{a}})\citenamefont {Hernandez}, \citenamefont
  {Nieves},\ and\ \citenamefont {Valverde}}]{Hernandez:2007qq}%
  \BibitemOpen
  \bibfield  {author} {\bibinfo {author} {\bibfnamefont {E.}~\bibnamefont
  {Hernandez}}, \bibinfo {author} {\bibfnamefont {J.}~\bibnamefont {Nieves}}, \
  and\ \bibinfo {author} {\bibfnamefont {M.}~\bibnamefont {Valverde}},\ }\href
  {\doibase 10.1103/PhysRevD.76.033005} {\bibfield  {journal} {\bibinfo
  {journal} {Phys. Rev.}\ }\textbf {\bibinfo {volume} {D76}},\ \bibinfo {pages}
  {033005} (\bibinfo {year} {2007}{\natexlab{a}})},\ \Eprint
  {http://arxiv.org/abs/hep-ph/0701149} {arXiv:hep-ph/0701149 [hep-ph]}
  \BibitemShut {NoStop}%
\bibitem [{\citenamefont {Tanabashi}\ \emph {et~al.}(2018)\citenamefont
  {Tanabashi} \emph {et~al.}}]{Tanabashi:2018oca}%
  \BibitemOpen
  \bibfield  {author} {\bibinfo {author} {\bibfnamefont {M.}~\bibnamefont
  {Tanabashi}} \emph {et~al.} (\bibinfo {collaboration} {Particle Data
  Group}),\ }\href {\doibase 10.1103/PhysRevD.98.030001} {\bibfield  {journal}
  {\bibinfo  {journal} {Phys. Rev.}\ }\textbf {\bibinfo {volume} {D98}},\
  \bibinfo {pages} {030001} (\bibinfo {year} {2018})}\BibitemShut {NoStop}%
\bibitem [{SAI()}]{SAID}%
  \BibitemOpen
  \href@noop {} {\enquote {\bibinfo {title} {{Institute for Nuclear Studies.
  The George Washington University Virginia Science and Technology Campus}},}\
  }\bibinfo {howpublished} {\url{http://gwdac.phys.gwu.edu/}},\ \bibinfo {note}
  {[Online; accessed 22-February-2019]}\BibitemShut {NoStop}%
\bibitem [{\citenamefont {Antonello}\ \emph {et~al.}(2015)\citenamefont
  {Antonello} \emph {et~al.}}]{Antonello:2015lea}%
  \BibitemOpen
  \bibfield  {author} {\bibinfo {author} {\bibfnamefont {M.}~\bibnamefont
  {Antonello}} \emph {et~al.} (\bibinfo {collaboration} {MicroBooNE, LAr1-ND,
  ICARUS-WA104}),\ }\href@noop {} {\  (\bibinfo {year} {2015})},\ \Eprint
  {http://arxiv.org/abs/1503.01520} {arXiv:1503.01520 [physics.ins-det]}
  \BibitemShut {NoStop}%
\bibitem [{\citenamefont {Sobczyk}\ \emph {et~al.}(2018)\citenamefont
  {Sobczyk}, \citenamefont {Hernández}, \citenamefont {Nakamura},
  \citenamefont {Nieves},\ and\ \citenamefont {Sato}}]{Sobczyk:2018ghy}%
  \BibitemOpen
  \bibfield  {author} {\bibinfo {author} {\bibfnamefont {J.}~\bibnamefont
  {Sobczyk}}, \bibinfo {author} {\bibfnamefont {E.}~\bibnamefont {Hernández}},
  \bibinfo {author} {\bibfnamefont {S.}~\bibnamefont {Nakamura}}, \bibinfo
  {author} {\bibfnamefont {J.}~\bibnamefont {Nieves}}, \ and\ \bibinfo {author}
  {\bibfnamefont {T.}~\bibnamefont {Sato}},\ }\href {\doibase
  10.1103/PhysRevD.98.073001} {\bibfield  {journal} {\bibinfo  {journal} {Phys.
  Rev.}\ }\textbf {\bibinfo {volume} {D98}},\ \bibinfo {pages} {073001}
  (\bibinfo {year} {2018})},\ \Eprint {http://arxiv.org/abs/1807.11281}
  {arXiv:1807.11281 [hep-ph]} \BibitemShut {NoStop}%
\bibitem [{\citenamefont {Hernandez}\ \emph
  {et~al.}(2007{\natexlab{b}})\citenamefont {Hernandez}, \citenamefont
  {Nieves},\ and\ \citenamefont {Valverde}}]{Hernandez:2006yg}%
  \BibitemOpen
  \bibfield  {author} {\bibinfo {author} {\bibfnamefont {E.}~\bibnamefont
  {Hernandez}}, \bibinfo {author} {\bibfnamefont {J.}~\bibnamefont {Nieves}}, \
  and\ \bibinfo {author} {\bibfnamefont {M.}~\bibnamefont {Valverde}},\ }\href
  {\doibase 10.1016/j.physletb.2007.02.051} {\bibfield  {journal} {\bibinfo
  {journal} {Phys. Lett.}\ }\textbf {\bibinfo {volume} {B647}},\ \bibinfo
  {pages} {452} (\bibinfo {year} {2007}{\natexlab{b}})},\ \Eprint
  {http://arxiv.org/abs/hep-ph/0608119} {arXiv:hep-ph/0608119 [hep-ph]}
  \BibitemShut {NoStop}%
\bibitem [{\citenamefont {Martin}\ and\ \citenamefont
  {Spearman}(1970)}]{Martin:102663}%
  \BibitemOpen
  \bibfield  {author} {\bibinfo {author} {\bibfnamefont {A.~D.}\ \bibnamefont
  {Martin}}\ and\ \bibinfo {author} {\bibfnamefont {T.~D.}\ \bibnamefont
  {Spearman}},\ }\href {https://cds.cern.ch/record/102663} {\emph {\bibinfo
  {title} {{Elementary-particle theory}}}}\ (\bibinfo  {publisher}
  {North-Holland},\ \bibinfo {address} {Amsterdam},\ \bibinfo {year}
  {1970})\BibitemShut {NoStop}%
\end{thebibliography}%
\end{document}